\def\numax{\nu_{\rm MAX}}
\def\numx{\nu_{\rm max}}
\def\msto{m_{s100}}
\def\rN{\rho_N}
\def\bbar{\overline B}
\def\ef{\varepsilon_{\rm fit}}
\def\al{\alpha_c}
\def\Pc{P_{\rm centr}}
\def\bunit{{\rm MeV\,fm^{-3}}}
\def\bmax{B_{\rm max}}
\def\Rms{R_{\rm ms}}
\def\ca{C_\alpha}

\documentclass{aa}
\usepackage{graphics}
\input{epsf}
\begin{document}
\thesaurus{06(02.04.1, 02.05.2, 08.14.1)}
\title{Strange stars - linear approximation of the EOS and
maximum QPO frequency}
\author{J.L. Zdunik}

\institute{N. Copernicus Astronomical Center, Polish
           Academy of Sciences, Bartycka 18, PL-00-716 Warszawa,
           Poland\\ e-mail:jlz@camk.edu.pl}
%
%\offprints{J.L. Zdunik}
%\mail{jlz@camk.edu.pl}
%
\date{Received / Accepted }

\maketitle
\begin{abstract}
In the present paper we study the equation of state of strange
matter build of the u,d,s quarks in the framework of the MIT bag
model. The scaling relations with bag constant are discussed and
applied to the determination of maximum frequency of a particle in
stable circular orbit around strange star. The highest QPO
frequency of 1.33~kHz observed so far is consistent with the
strange stars models for which the maximum QPO frequency is
1.7--2.4~kHz depending on the strange quark mass and the QCD
coupling constant. The linear approximation of the equation of
state is found and the parameters of this EOS are determined as a
functions of strange quark mass,  QCD coupling constant and bag
constant. This approximation reproduces exact results within an
error of the order of 1\% and can be used for the complete study
of the properties of strange stars including microscopic stability
of strange matter and determination of the total baryon number of
the star.


\end{abstract}

\section{Introduction}

The idea of the compact stars build of strange matter was
presented by Witten (\cite{witten}) and the models of
stars were calculated using various models of strange
matter by Haensel et al. (\cite{hzs}) and
Alcock et al. (\cite{afo}).
The main idea is that the u,d,s matter is ground state of matter
at zero pressure (self-bound strange quark
 matter) i.e.:
\begin{equation}
\mu_0\equiv\mu(P=0) < M({\rm ^{56}Fe})=930.4~{\rm MeV}
\label{udsstab}
\end{equation}

There were two aspects of rather extensive studies of strange stars
properties. The first one in the context of maximum rotational velocity
of the star (Frieman \& Olinto \cite{fo}, Glendenning \cite{glen89},
Prakash et al. \cite{Prakash}, Zdunik \& Haensel \cite{zh90})
 was related to the announcement
reporting the detection of the half millisecond pulsar in 1989 (which after one
year turned out to be erroneous). This observation stimulated detailed
investigations of the limits on the rotational frequency of dense stars
 which excluded nearly all neutron stars models leaving strange stars
as a possible explanation.

At present the increasing interest in strange stars is a result of
QPOs observations in low-mass X-ray binaries
LMBX and some difficulties with the
explanations of its properties by neutron star models under the
assumption that QPOs represent Keplerian frequencies of the particles in
an accretion disk (Kaaret et al. 1997, Klu{\'z}niak 1998, Zhang et al. 1998,
 Miller et al. 1998, Thampan et al. 1999, Schaab \& Weigel 1999).
In this paper the maximum values of these frequencies are found
for a broad set of the parameters describing a strange matter EOS.
Strange star models are consistent with the maximum observed QPO
frequency $1.33$~kHz in 4U 0614+09 (van Straaten et al.
\cite{Straaten}).

Frieman \& Olinto (\cite{fo}) mentioned the approximation of the
EOS of strange matter by the linear function (see also Prakash et
al. \cite{Prakash}) . I present here the linear form of the EOS of
strange matter with parameters expressed as a polynomial functions
of the physical constants describing MIT bag model (strange quark
mass and QCD coupling constant). These approximate formulae allow
us to write down the pressure vs. density dependence in a simple
form $P=a\cdot(\rho-\rho_0)$. The consequence of this form are
scaling laws of all stellar quantities with appropriate powers of
$\rho_0$. This linear EOS is complete in the sense that it
contains not only pressure and energy density, which enter the
relativistic stress-energy tensor and are sufficient for the
determination of the main parameters of the star (mass, radius),
but also the formula for baryon number density (or equivalently
baryon chemical potential), necessary in the microscopic stability
considerations and determination of the total baryon number of the
star.

\section{Strange stars and maximum QPO frequency}

We describe the quark matter using the phenomenological MIT bag
model (see, e.g., Baym, \cite{baym}). The quark matter is the
mixture of the massless $u$ and $d$ quarks, electrons and massive
$s$ quarks. The model is described in detail in Farhi \& Jaffe (\cite{fj}),
where the formulae for physical parameters of a strange matter are also presented.
There are the following physical quantities entering this model:
$B$ -- the bag constant, $\al$ -- the QCD coupling constant and $m_s$ --
the mass of the strange quark. It is necessary to introduce also
the parameter $\rN$ -- the renormalization  point. Following Farhi
\& Jaffe (\cite{fj}) we choose $\rN=313$~MeV.

The consequence of this model of strange matter is scaling of all
thermodynamic functions with some powers of $B$.
Knowing the EOS for given $\al$, $m_s$  and $B_0$ we can obtain thermodynamic
quantities for other value $B$ from the following formulae:
\begin{equation}
\begin{array}{lclclcl}
P_{[B]}&=&P_{[B_0]}\cdot(B/B_0),&\rho_{[B]}&=&\rho_{[B_0]}\cdot(B/B_0),\\
n_{[B]}&=&n_{[B_0]}\cdot(B/B_0)^{3/4},&\mu_{[B]}&=&\mu_{[B_0]}\cdot(B/B_0)^{1/4}
\end{array}
\label{scleos}
\end{equation}
where the resulting EOS for $B$ is determined for the same value of $\al$ but for
the strange quark mass given by the relation:
\begin{equation}
%\begin{array}{lcl}
m_s(B)=m_s(B_0)\cdot (B/B_0)^{1/4}\label{sclms}
\end{equation}

The advantage of this approximations is the elimination of one
parameter ($B$) from the calculations of the EOS. The dependence of
all parameters on $B$ is very well defined and one can take this
into account using simple scaling laws. It is therefore sufficient to study
the EOS in two parameter space: $\al$ and $m_s$
for chosen value of $B_0$ and then obtain results for other $B$ from
Eqs (\ref{scleos}).

\begin{figure}     %1
\resizebox{\hsize}{!}
{\includegraphics{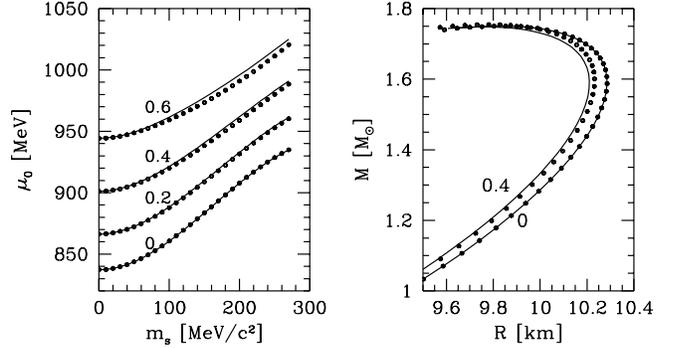}}
\caption[]{The comparison between exact values of $\mu_0$, $M$ and
$R$ obtained by the calculation of the EOS for $B=60~\bunit$ and
$B=90~\bunit$ and scaling relations. The values for $B=90$ was
rescaled to the value of $B_0=60$ using the relations
$\mu\to\mu\cdot(B/B_0)^{-1/4}$, $m_s\to m_s\cdot(B/B_0)^{-1/4}$,
$M\to M\cdot(B/B_0)^{1/2}$ and $R\to R\cdot(B/B_0)^{1/2}$. In the
case of {\it exact} scaling relations the results for $B=60$ and
rescaled results for $B=90$ would be the same.} \label{miubcmp}
\end{figure}

Strictly speaking the scaling relations (\ref{scleos}) hold exactly
for the parameter $\rN$ rescaled by the relation
$\rN(B)=\rN(B_0)\cdot (B/B_0)^{1/4}$.
However the renormalization point is fixed and this is the
source of small discrepancy between the true results for different
values of bag constant and the scaling relations presented in this
paper. This difference is presented in Fig. \ref{miubcmp} in the
case of the baryon chemical potential at zero pressure
calculated directly and using
scaling relation (\ref{scleos}). We choose $\mu_0$ because
this thermodynamic function is crucial in microscopic stability considerations.
The models of strange quark matter correspond
to two values of bag constant: $B_0=60~{\rm MeV\,fm^{-3}}$
and $B=90~{\rm MeV\, fm^{-3}}$.
The strange quark masses
in this two cases fulfill the relation (\ref{sclms}), namely
$m_s(60)=200~{\rm MeV}$ and $m_s(90)=221.33~{\rm MeV}$.

We see that scaling formulae give exact results for $\al=0$
because then $\rN$ does not enter the EOS. The maximum error of
the scaling formulae (\ref{scleos}) is less than 0.5\%.

 From the microscopic point of view the strange quark matter is unstable
at zero pressure for some values of $\al$ and $m_s$ discussed in this paper
 (large $\al$ and $m_s$), i.e. equation of state does not
fulfill the relation (\ref{udsstab}). But from presented results
we can obtain, using scaling formulae, the configurations which
correspond to smaller value of $B$ and are microscopically stable.
The maximum value of $B$ for which the strange quark matter is
stable can be obtained from the equation:
\begin{equation}
 \mu_0[B_0,\al,m_s(\bmax/B_0)^{-1/4}]\,
\left({\bmax\over B_0}\right)^{1/4}=930.4~{\rm MeV}
\label{bmax}
\end{equation}

%%%%%%%%%%%%%%%%%
%%%%%%%%%%%%%%%%%%%%%%%%
\medskip

The strange star configurations are calculated by solving
Oppenheimer-Volkoff equations in the case of spherical symmetry.
For given central pressure $\Pc$ we obtain stellar parameters
such as gravitational mass $M$, baryon number $A$,
moment of inertia for slow rigid rotation
$I$ and gravitational surface redshift $z$.
All these quantities are subject to the scaling
formulae  similar to those describing the
EOS (Eq. \ref{scleos}). If we calculate the star for $B_0$  with
central pressure equal to $\Pc[B_0]$ we know all parameters of the
corresponding star with $\Pc=\Pc[B_0]\,B/B_0$ in the model with
bag constant $B$ (cf. Witten \cite{witten}, Haensel et al. \cite{hzs}).
In general for the stellar parameter $X$ the following equality holds:
\begin{equation}
X[B,\al,m_s]=X[B_0,\al,m_s(B/B_0)^{-1/4}]\,(B/B_0)^{-k}
\label{sclpar}
\end{equation}
where $k=1/2$ in the case of mass and radius ($X=R, M$), $k=3/4$ for $X=A$,
$k=3/2$ for $I$ and $k=0$ for $z$.

In Fig. \ref{miubcmp} we see that scaling formulae (\ref{sclpar}) reproduce
the stellar parameters $M$ vs. $R$ with an error less than 0.3\%.

\begin{figure}     %
\resizebox{\hsize}{!}
{\rotatebox{-90}{\includegraphics{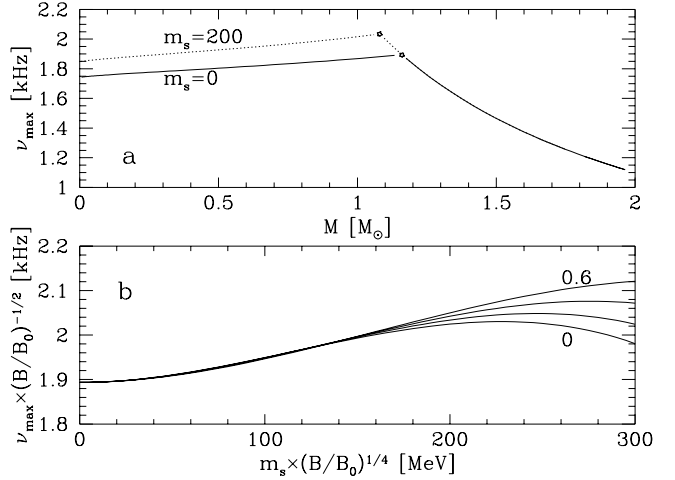}}}
\caption[]{The value of the frequency of a particle
in the innermost stable circular orbit around nonrotating strange star for
strange matter models with $m_s=0$ and $m_s=200~{\rm MeV}$ (panel {\bf a})
 and the maxima of those functions as a function
of $m_s$ for $\al=0,~0.2,~0.4, ~0.6$ (panel {\bf b}).
The value of bag constant is fixed and equal to $B_0=60~\bunit$ }
\label{vmaxm}
\end{figure}

These scaling laws so far have been mainly
exploited in the case
of maximum mass of the star (or equivalently maximum rotational frequency).
Of course these relations refer to all stellar configurations (e.g. the
curve $M(R)$) and as an interesting example we can consider the point
defined by the equality:
\begin{equation}
R=\Rms=3R_g=6{GM\over c^2} \label{radms}
\end{equation}
which corresponds to the maximum possible frequency of a particle
in stable circular orbit $\numx$ around nonrotating star. Because
the scaling properties of $R$ and $M$ are the same ($\sim
B^{-1/2}$) one can solve the Eq. \ref{radms} independently of $B$.
For stellar masses higher than the solution of the Eq. \ref{radms}
the maximum Keplerian frequency of an orbiting particle is defined
by the marginally stable orbit at $\Rms$ and for lower by the
stellar radius $R$ (Fig. \ref{vmaxm}a). In both cases this
frequency scales as $B^{1/2}$  and for other
values of $B$ the patterns of Fig. \ref{vmaxm}a
do not change, provided one rescales the axes and $m_s$ (Eqs
\ref{sclms}, \ref{sclpar}, \ref{vscl}).

\begin{figure}     %
\resizebox{\hsize}{!}
{\rotatebox{-90}{\includegraphics{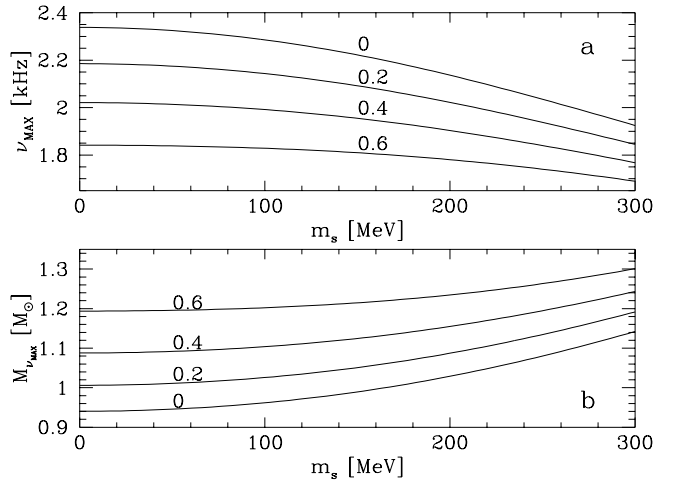}}}
\caption[]{The absolute maximum  value of the frequency of a
particle in stable circular orbit around nonrotating strange star
$\numax$ obtained for $B=\bmax(m_s,\al)$ (Eq. \ref{bmax}) (panel {\bf a})
and the corresponding values of stellar mass $M$ -- the minimum
mass for which the existence of the marginally stable orbit is
possible (panel {\bf b}).}
\label{vmaxms}
\end{figure}

The maximum values of $\nu$ as a function of strange matter
parameters are presented in Fig. \ref{vmaxm}b for $B_0=60~\bunit$.
For other values of $B$ these results scale as $B^{1/2}$ i.e.:
\begin{equation}
\numx[B,\al,m_s]=\numx[B_0,\al,m_s(B/B_0)^{-1/4}]\,
\left({B\over B_0}\right)^{1/2}
\label{vscl}
\end{equation}
which allows us to determine the absolute maximum value of
$\numx$ consistent with the requirement of the stability of strange
matter at zero pressure (Eq. \ref{bmax}) by putting the maximum
allowed value of $B$ into Eq. (\ref{vscl}). The result of this
procedure $\numax\equiv\numx(\bmax)$ is presented in Fig. \ref{vmaxms}a.
We see that this limit is well above the highest observed QPO frequency
$\nu_{\rm obs}=1.33~{\rm kHz}$
recently reported by van Straaten et al. (\cite{Straaten}).
Thus at present the observations of the highest QPO
frequency do not constrain strange matter models unless we do not
make the assumptions about the mass of the star in LMBX.
To set some limits on strange matter parameters the QPOs must be observed
at frequencies larger than $1.8$~kHz.
Our conclusion is true also for moderate rotation rates of strange
star due to the initial increase of the frequency at marginally
stable orbit which in the case of stars with mass slightly above
the solution of Eq. \ref{radms} results in $\numx$ very close to
the value corresponding to the nonrotating stars (Zdunik et al.
\cite{zdunik00}, Datta et al. \cite{datta00}).

It should be mentioned that in principle analysis of the observational
data can strongly support the existence of the marginally stable orbit.
Such a conclusion have been recently presented by Kaaret et al.
(\cite{Kaaret99}) in the case of  4U~1820-30. This interpretation
is equivalent to the condition $R<R_{ms}$ setting the lower bound for
the mass of the star. This minimum mass limits are presented in
Fig. \ref{vmaxms}b as a function of $m_s$ and $\al$. Below this mass
the innermost stable orbit is located at the surface of the star.

%%%%%%%%%%%%%%%%%%%%%%%%%%%%%%%%%%%%%%%%%%%%%%%%%%%%%%%%%%%%%%%%%%%%%%%%%%%%%%
%%%%%%%%%%%%%%%%%%%%%%%%%%%%%%%%%%%%%%%%%%%%%%%%%%%%%%%%%%%%%%%%%%%%%%%%%%%%%%

%%%%%%%%%%%%%%%%%%%%%%%%%%%%%%%%%%%%%%%%%%%%%%%%%%%%%%%%%%%%%%%%%%%%%%%%%%%%%%
%%%%%%%%%%%%%%%%%%%%%%%%%%%%%%%%%%%%%%%%%%%%%%%%%%%%%%%%%%%%%%%%%%%%%%%%%%%%%%

%%%%%%%%%%%%%%%%%%%%%%%%%%%%%%%%%%%%%%%%%%%%%%%%%%%%%%%%%%%%%%%%%%
\section{Linear interpolation of EOS}

\begin{figure}     %2
\resizebox{\hsize}{!}
{\includegraphics{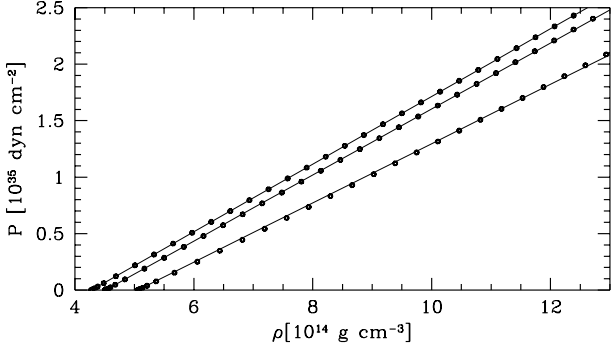}}
\caption[]{The linear approximation of the EOS for strange quark
matter for three choices of $(m_s, \al)$: $(0,0)$, $(100,0.4)$,
$(250,0.6)$ (from left to right). The points represent exact
results obtained by the numerical calculation of the EOS for
$B=60~\bunit$ and lines correspond to the approximate formula
\ref{lineos}.} \label{eos}
\end{figure}

In the present section we discuss the interpolation of equation of
state of strange quark matter by the linear function of $\rho$.
For a broad set of parameters $\al$ and $m_s$ the dependence
$P(\rho)$ can be very well approximated by the linear function,
being exactly linear at the $m_s=0$ limit. The EOS for strange
quark matter is presented in Fig. \ref{eos}.

The main parameters of the matter are described by the
following formulae:
\begin{equation}
\begin{array}{lcl}
P(\rho)&=&{1\over3}(1+\ef)(\rho-\rho_0)c^2\\
n(P)&=&n_0\cdot\left[1+\left(4-\displaystyle{3\ef\over
1+\ef}\right)\displaystyle{P\over\rho_0c^2}\right]^{3/(4+\ef)}
\end{array}
\label{lineos}
\end{equation}
where $\rho_0$ and $n_0$ are energy and number density of the
strange quark matter at zero pressure.
The second equation is implied by the first law of thermodynamics.

The equations (\ref{lineos}) allow us to determine all
thermodynamic quantities characterizing strange quark matter in
linear interpolation of the EOS, e.g. the baryon chemical potential:
\begin{equation}
\mu(P)=\mu_0
\left(1+{4+\ef\over 1+\ef}{P\over \rho_0 c^2}\right)^{(1+\ef)/(4+\ef)}
\label{mulineos}
\end{equation}
where $\mu_0=\rho_0 c^2/n_0$.

Because for all thermodynamic quantities the scaling relations
with $B$ hold we can calculate all parameters needed for the
linear EOS for chosen value of $B$ and then rescale them using
equations (\ref{scleos}). Thus the main point is to determine
three parameters entering equations (\ref{lineos}), as a functions
of $\al$ and $m_s$: $\rho_0(\al,m_s)$, $n_0(\al,m_s)$ and
$\ef(\al,m_s)$. These functions can be very well approximated by
the polynomial of $m_s$ with coefficients depending on the value
of $\al$. The formulae obtained by the least-squares fit to the
exact results are following:
\begin{eqnarray}
n_0&=&(a^n_0+a^n_2\msto^2+a^n_3\msto^3){\bbar}^{3/4}\label{interpln}\\
&a^n_0&=0.28660\ca^{1/4}\nonumber\\
&a^n_2&=(0.010788+0.0032046\al)/{\bbar}^{1/2}\nonumber\\
&a^n_3&=-0.0044248\sqrt{\ca}/{\bbar}^{3/4}\nonumber\\
\mu_0&=&(a^{\mu}_0+a^{\mu}_2\msto^2+a^{\mu}_3\msto^3){\bbar}^{1/4}\label{interplmu}\\
&a^{\mu}_0&=837.260/\ca^{1/4}\nonumber\\
&a^{\mu}_2&=(46.616-16.848/\ca)/{\bbar}^{1/2}\nonumber\\
&a^{\mu}_3&=(-10.482+4.5211/\ca)/{\bbar}^{3/4}\nonumber\\
\ef&=&a^{\epsilon}_2\msto^2+a^{\epsilon}_3\msto^3\label{interple}\\
&a^{\epsilon}_2&=(-0.035745+0.013366\al+0.023246\al^2)/{\bbar}^{1/2}\nonumber\\
&a^{\epsilon}_3&=(0.0055488-0.0054232\al-0.0069193\al^2)/{\bbar}^{3/4}\nonumber
\end{eqnarray}
where $\msto$ is strange quark mass in units 100 MeV i.e.
$\msto=m_sc^2\,[{\rm MeV}]/100$, $\bbar=B\,[\bunit]/60$ and
$\ca=1-2\al/\pi$.

\begin{figure}     %
\resizebox{\hsize}{!}
{{\includegraphics{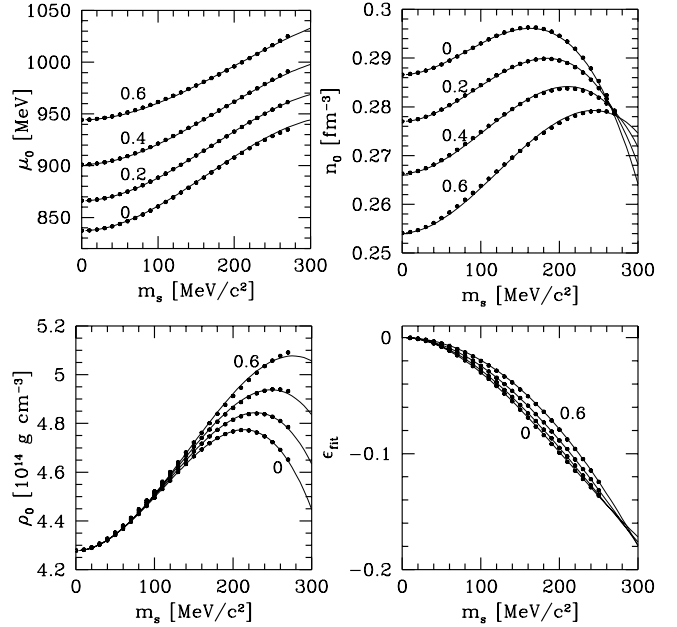}}}
\caption[]{The comparison between exact values of the main
parameters of the strange matter (baryon chemical potential,
particle number density, mass-energy density) at zero pressure and
the parameter $\ef$ entering the linear approximate EOS of strange
quark matter $P=1/3(1+\ef)(\rho-\rho_0)$ obtained by the
interpolation of the EOS for $B=60~\bunit$ (points) and
approximate formulae (\ref{interpln}-\ref{interple}) (lines).}
\label{pareos0}
\end{figure}

The energy-matter density at zero pressure is equal to
$\rho_0=n_0\mu_0/c^2$.
The functions $\rho_0(m_s)$ and  $\ef(m_s)$ entering directly Eq. \ref{lineos}
and $\mu_0(m_s)$  for chosen
values of $\al$ and accuracy of the above interpolations are
presented in Fig.~\ref{pareos0}.

\begin{figure}     
\resizebox{\hsize}{!}
{{\includegraphics{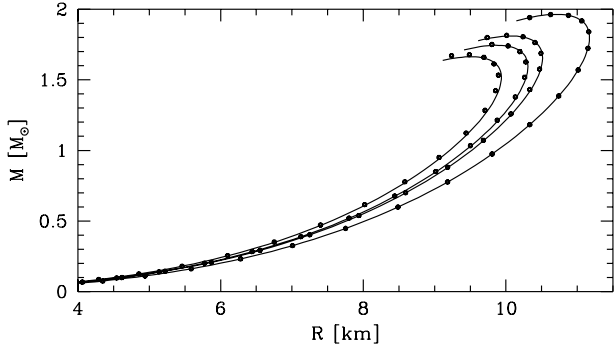}}}
\caption[]{The
mass-radius relations for strange stars calculated using
exact EOS (points) and the linear approximation of the EOS
 $P\sim(\rho-\rho_0)$ (lines) for $B=60~\bunit$. The curves correspond
to four different parameters of strange matter $(m_s,\al)$: $(0,0)$,
$ (150,0.3)$, $(200,0)$, $(250,0.6)$ (from right to left).}
\label{mrlincmp}
\end{figure}

One can test the accuracy of the interpolation of the EOS by the
linear function calculating the stellar configurations build of
such a matter and comparing results with stars calculated using
exact equation of state. Such a comparison is presented in Fig.
\ref{mrlincmp}. The relative error in this procedure is always less than
1\% in the case of mass, radius and moment of inertia.

Except for the case of strange matter with massless quarks, 
studied in detail in the literature, the linear
form of the equation of state of self-bound matter have been
used  so far mainly for the stellar mass and radius
determination (e.g. Haensel \& Zdunik \cite{hz89}, Lattimer et al.
\cite{latt90}). The authors exploited the relation
$P=a(\rho-\rho_0)$, because the quantities sufficient in these
considerations are pressure and energy density, which enter the
stress-energy tensor and TOV equations of hydrostatic equilibrium.
However for the complete study of the properties of star the full
microscopic description of the matter is needed, including baryon
chemical potential and particle number density. In our EOS the
formula for $n(P)$ (Eq. \ref{lineos}) enables us to determine the
total baryon number of the stellar configuration and use this
model of strange matter, for example, for the discussion of the
conversion of neutron stars into strange stars (Olinto \cite{olinto87},
Cheng \& Dai \cite{Cheng96}, Bombaci \& Datta \cite{bd00}). During this
process the total baryon number of the star is fixed. Using the
relation (\ref{interplmu}) one can find the ranges of the values
of $m_s$, $\al$ and $B$ consistent with the requirement that
strange matter is ground state of matter at zero pressure (Eq.
\ref{udsstab}). As a result the allowable parameters of strange
stars (e.g. $M$, $\numax$) can be found without complicated and
time-consuming calculations of the microscopic properties of
strange matter.

\section{Discussion and conclusions}

In the present paper I analyzed the accuracy of the scaling
properties of strange matter parameters with the value of the bag
constant $B$ in the framework of MIT bag model of quark matter.
The scaling formulae reproduce the main stellar parameters with an
error less than 1\%. This allow us to use them to determine the
maximum possible frequency of a particle in stable circular orbit
around strange star. The absolute maximum QPO frequency that can
be accommodated by strange stars ranges from 1.7 to 2.4~kHz
depending on the values of a strange matter parameters: strange
quark mass and QCD coupling constant. Thus the present status of
observational data ($\nu_{\rm max}^{\rm obs}=1.33$~kHz) cannot
exclude strange stars as source of QPOs in LMBX. The frequencies
of QPO in the very high range (larger than $1.8$~kHz), if
observed, may set some bounds on parameters of strange matter,
excluding large $m_s$ and large $\al$. We should mention that such
a high values of $\nu_{\rm QPO}$ cannot be understood in terms of
neutron stars (Thampan et al. \cite{Thampan99}).

The high accuracy of the simple scaling laws with $B$ is strictly
connected with the possibility of the approximation of the
equation of state by the linear function $P=a(\rho-\rho_0)$ for
which such a scaling properties are well known and exact. For this
linear EOS the parameters of the strange stars scale with the
powers of $\rho_0$ for fixed value of $a$ (analogous to the
scaling laws with $B$, Eqs. \ref{sclpar}, \ref{vscl}). It should
be stressed that these scaling properties are valid not only for
static stellar configurations but also in some dynamical problems
(e.g. parameters of the rotating star, Gourgoulhon et al.
\cite{ghlp}). For fixed $\rho_0$ one can apply also the
approximate scalings of the stellar parameters at the maximum mass
point with appropriate powers of $a$ (Lattimer et al.,
\cite{latt90}) which have been recently confirmed by Stergioulas
et al. (\cite{skb}) for stars rotating at Keplerian frequency.

The presented linear approximation of the equation of state allow
us to determine the dependence of many properties of strange stars
on the physical parameters of a matter ($m_s$, $\al$) using very
simple form of the EOS. For a broad set of  $m_s$ and $\al$ the
values of $\ef$ and $\rho_0$ entering linear EOS can be very
accurately obtained by a polynomial formulae. In particular the
formula for the baryon chemical potential enables us to study the
microscopic stability of strange matter and make a complete
discussion of the resulting constraints on strange stars
parameters. The expression for baryon number density is necessary
in the consideration of the conversions of neutron stars into
strange stars with total baryon number conserved.

It is worth noticing that the linear approximation of $P(\rho)$
dependence (Eq. \ref{lineos}) can be used also for other models of
strange matter, self-bound at high density $\rho_0$. The
expressions for $n(P)$ and $\mu(P)$ allow us to determine all
microscopic properties of the matter given the values of $\rho_0$
and $n_0$.

\begin{acknowledgements}
 This research was partially supported by the KBN grant No. 2P03D.014.13.
I am very grateful to P.~Haensel for careful reading of manuscript and helpful
comments and suggestions.
\end{acknowledgements}

\end{document}